\documentclass{llncs}

\newif\ifdraft
\draftfalse

\usepackage[english]{babel}
\usepackage[T1]{fontenc}
\usepackage[utf8]{inputenc}
\usepackage{lmodern}
\usepackage{textcomp}
\ifdraft
\usepackage[show]{ed/ed}
\usepackage[eso-foot,today]{svninfo}
\svnInfo $Id: gencs-lod.tex 1628 2010-03-10 21:21:22Z clange $
\svnKeyword $HeadURL: https://svn.omdoc.org/repos/jomdoc/doc/pubs/eswc-demo10/gencs-lod.tex $
\else
\usepackage{microtype}
\usepackage[hide]{ed/ed}
\fi

\usepackage[hyperref=auto,firstinits=true]{biblatex}
\bibliography{kwarc}
\usepackage[babel]{csquotes}
\MakeAutoQuote{“}{”}
\MakeAutoQuote*{‘}{’}
\usepackage{clange}
\usepackage{stex-logo}
\usepackage{paralist}
\usepackage{graphicx}
\usepackage{wrapfig}
\usepackage{soul}
\setul{.25ex}{}

\usepackage{url}

\def\sys{\textsc{Planetary}\xspace}
\def\infobar{\texttt{InfoBar}\xspace}
\def\foldingbar{\texttt{FoldingBar}\xspace}

\title{The Planetary System: Executable \ul{S}cience, \ul{T}echnology, \ul{E}ngineering and \ul{M}ath Papers}
\author{Christoph Lange \and Michael Kohlhase \and Catalin David \and Deyan Ginev 
\and Andrea Kohlhase
 \and Bogdan Matican \and Stefan Mirea \and Vyacheslav Zholudev}
\institute{Computer Science, Jacobs University Bremen, Germany  \email{\{ch.lange,m.kohlhase,c.david,d.ginev,a.kohlhase,
b.matican,s.mirea,v.zholudev\}@jacobs-university.de}}
\begin{document}

\maketitle

\begin{abstract}
  Executable scientific papers contain not just layouted text for reading.  They contain, or link to, machine-comprehensible representations of the scientific findings or experiments they describe.  Client-side players can thus enable readers to “check, manipulate and explore the result space” \cite{Elsevier:EPC11}.  We have realized executable papers in the STEM domain with the \sys system.  Semantic annotations associate the papers with a content commons holding the background ontology, the annotations are exposed as Linked Data, and a frontend player application hooks modular interactive services into the semantic annotations.
\end{abstract}

\section{Application Context: STEM Document Collections}
\label{sec:application}

The \sys system~\cite{DGKC:eMath30} is a semantic social environment for document collections in Science, Technology, Engineering and Mathematics (STEM).  STEM documents have in common that they describe concepts using mathematical formulæ, which are composed from mathematical symbols – operators, functions, etc.\@ –, which have again been defined as more foundational mathematical concepts in mathematical documents.  Thus, there is a dynamically growing ontology of domain knowledge.  The domain knowledge is structured along the following, largely independent dimensions \cite{KohKohLan:difcsmse10:biblatex,Lange:OntoLangMathSemWeb}: \begin{inparaenum}[(i)]\item logical and functional structures, \item narrative and rhetorical document structures, \item information on how to present all of the former to the reader (such as the notation of mathematical symbols), \item application-specific structures (\eg for physics), \item administrative metadata, and \item users' discussions about artifacts of domain knowledge\end{inparaenum}.

We have set up \sys instances for the following paradigmatic document collections: \begin{inparaenum}[(i)]\item a browser for the ePrint \arxiv~\cite{arXMLiv-frontend:online}, \item a reincarnation of the PlanetMath mathematical encyclopledia~\cite{planetmathredux:on} (where the name {\sys} comes from), \item a companion site to the general computer science (GenCS) lecture of the second author~\cite{PlanetGenCS:on,DKLRZ:PubMathLectNotLinkedData10}, and \item an atlas of theories of formal logic~\cite{PlanetLATIN:on}\end{inparaenum}.  This list is ordered by increasing machine-comprehensibility of the representation and thus, as explained below, by increasing “executability” of the respective papers.  All instances support browsing and fine-grained discussion.  The \planetmath and GenCS collections are editable, as in a wiki\footnote{\sys reuses technology of our earlier semantic wiki \swim \cite{lange:swim-demo08}.}, whereas the \arxiv and Logic Atlas corpora have been imported from external sources and are presented read-only.  We have prepared demos of selected services in all of these instances.

\section{Key Technology: Semantics-Preserving Transformations}
\label{sec:technology}

Documents published in \sys become flexible, adaptive interfaces to a {\emph{content commons}} of domain objects, context, and their relations.  This is achieved by providing an integrated user experience through a set of interactions with documents based on an extensible set of client- and server side services that draw on explicit (and thus machine-understandable) representations in the content commons.  We have implemented or reused ontologies for all structures of STEM knowledge (\cite{Lange:OntoLangMathSemWeb} gives an overview).  Annotations of papers with terms from these ontologies act as hooks for local interactive services.  By translation, \sys makes the structural ontologies editable in the same way as the papers, so that the community can adapt and extend them to their needs.

The sources of the papers are maintained in {\LaTeX} or the semantic mathematical markup language {\omdoc}~\cite{Kohlhase:omdoc1.2}.  For querying and information retrieval, and interlinking with external knowledge – including discussions about concepts in the papers, but also remote Linked Datasets –, we extract their semantic structural outlines to an \rdf representation, which is accessible to external services via a \sparql endpoint and as Linked Data \cite{DKLRZ:PubMathLectNotLinkedData10}.  For human-comprehensible presentation, we transform the sources to {\xhtml}+{\mathml}+\linebreak[1]{\svg} \cite{DKLRZ:PubMathLectNotLinkedData10}.  These papers gain their “executability” from embedded semantic annotations: Content {\mathml}\footnote{or the semantically equivalent \openmath \cite{BusCapCar:2oms04}} embedded into  formulæ \cite{CarlisleEd:MathML10}, and an \rdfa subgraph of the above-mentioned \rdf representation embedded into {\xhtml} and {\svg}.

The amount of semantic annotations depends on the source representation:
\begin{inparaenum}[(i)]
\item The \arxiv corpus – 500+K scientific publications – has {\LaTeX} sources, most of which merely make the section structure of a document machine-comprehensible, but hardly the fine-grained functional structures of mathematical formulæ, statements (definition, axiom, theorem, proof, etc.), and theories.  We have transformed the papers to {\xhtml}+{\mathml}, preserving semantic properties like formula and document structure \cite{arXMLiv-frontend:online}.
\item The \planetmath corpus is maintained inside \sys; it additionally features subject classification metadata and semi-automatically annotated concept links~\cite{GKX:NNexusAutoLinker09}, which we preserve as \rdfa.
\item The GenCS corpus is maintained in {\sTeX}, a semantics-extended {\LaTeX}~\cite{KohKohLan:ssffld10:biblatex}, inside \sys.  {\sTeX} makes explicit the functional structures of formulæ, statements, and theories, narrative and rhetorical structures, information on notation, as well as – via an \rdfa-like extensibility – arbitrary administrative and application-specific metadata.  This structural markup is preserved as Content \mathml and \rdfa in the human-comprehensible output.  In this translation, \omdoc, an \xml language semantically equivalent to {\sTeX}, serves as an intermediate representation.
\item The Logic Atlas is imported into \sys from an external \omdoc source but otherwise treated analogously to the GenCS corpus.
\end{inparaenum}

\section{Demo: Interactive Services and the \sys \api}
\label{sec:demo}

\begin{figure}%
\begin{minipage}{.53\textwidth}
\includegraphics[width=\textwidth]{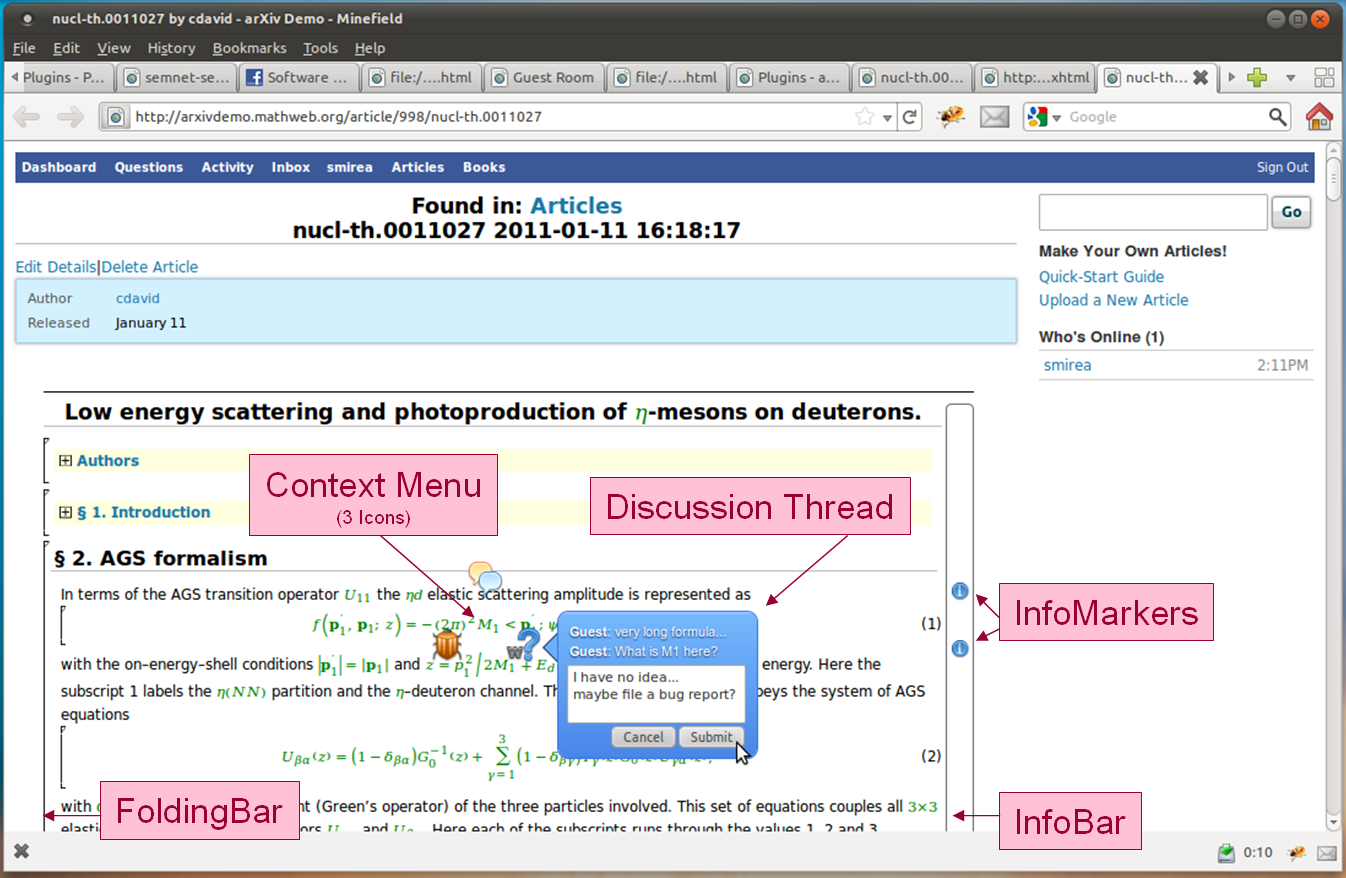}
\end{minipage}\hspace{.09\textwidth}%
\begin{minipage}{.38\textwidth}
  \begin{tabular}{c}
    \fbox{\includegraphics[width=\textwidth]{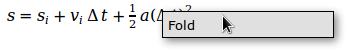}}\\
    $\Downarrow$\\
    \fbox{\includegraphics[width=\textwidth]{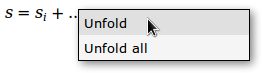}}\\
  \end{tabular}
\end{minipage}
\caption{Interacting with an \arxiv article via \foldingbar, \infobar, and localized discussions.  On the right: localized folding inside formulæ}\label{fig:infobar}
\end{figure}
% \begin{wrapfigure}r{2cm}
%   \vspace*{-6ex}
%   \strut\hspace*{-.5em}\strut\begin{tabular}{|c|}\hline
%     \includegraphics[width=2cm]{images/fold}\\\hline
%     \includegraphics[width=2cm]{images/folded}\\\hline
%   \end{tabular}\vspace*{-5ex}
% \end{wrapfigure}
\looseness=-1
\noindent Our demo focuses on how \sys makes \stem papers executable – by hooking interactive services into the annotations that the semantics-preserving translations put into the human-comprehensible presentations of the papers.  Services are accessible locally via a context menu for each object with (fine-grained) semantic annotations – \eg a subterm of a formula –, or via the “{\infobar}”, as shown in fig.~\ref{fig:infobar}.  The menu has one entry per service available in the current context; the \infobar indicates the services available for the information objects in each line of the paper.  In the image on the right of fig.~\ref{fig:infobar}, we selected a subterm and requested to fold it, \ie to simplify its display by replacing it with an ellipsis.  The \foldingbar on the left, similar to source code IDEs, enables folding document structures, and the \infobar icons on the right indicate the availability of local discussions.  Clicking them highlights all items with discussions; clicking any of them yields an icon menu as shown in the center.  The icon menu for the discussion service allows for reporting problems or asking questions using a \stem-specifically extended argumentation ontology~\cite{LBGBH08:SIOC-argumentation}.
\begin{figure}\centering
  \begin{tabular}{|p{7cm}|p{5cm}|}\hline 
  %\vspace{-7cm}
  %\fbox{\includegraphics[width=2.5cm]{images/iconmenu-GenCS}}\newline\vspace{2cm}
  \vspace{-5.5cm}\hspace{2cm}\fbox{\includegraphics[width=2cm]{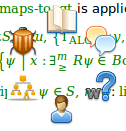}}\newline\vspace{-.25cm}\newline\includegraphics[width=5.5cm]{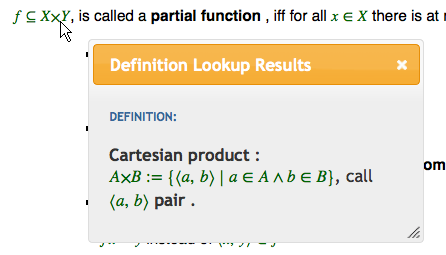} & 
  \includegraphics[width=5cm]{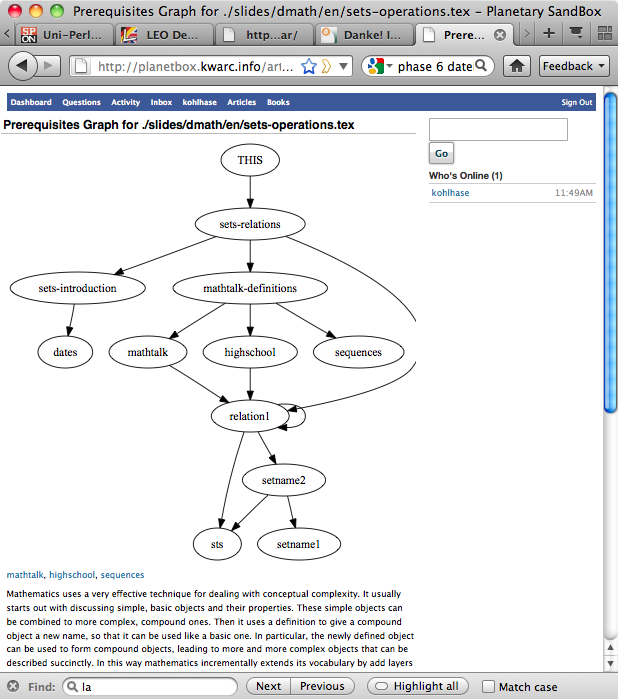}\\\hline
\end{tabular}
\caption{Definition Lookup and Prerequisites Navigation}\label{fig:definition-lookup}
\end{figure}
The richer semantic markup of the GenCS and Logic Atlas collections enable services that utilize  logical and functional structures – reflected by a different icon menu.  Fig.~\ref{fig:definition-lookup} demonstrates looking up a definition and exploring the prerequisites of a concept.  The definition lookup service obtains the \uri of a symbol from the annotation of a formula and queries the server for the corresponding definition.  The server-side part of the prerequisite navigation service obtains the transitive closure of all dependencies of a given item and returns them as an annotated {\svg} graph.  Computational services make mathematical formulæ truly executable: The user can send a selected expression to a computer algebra web service for evaluation or graphing \cite{DLR:InteractDocCAS-JOBAD-Alpha10}, or have unit conversions applied to measurable quantities \cite{GLR:WebSvcActMathDoc09}.  Finally, besides these existing services, we will demonstrate the ease of realizing additional services – within the \sys environment or externally of it.  The \api for services running as scripts in client-side documents is essentially defined by the in-document annotations, the underlying structural ontologies that are retrievable from the content commons, the possibility to execute queries against the content commons, and the extensibility of the client-side user interface.

\section{Related Work}
\label{sec:research}

Like a \textbf{semantic wiki}, \sys supports editing and discussing resources.  Many wikis support {\LaTeX} formulæ, but without fine-grained semantic annotation.  They can merely \emph{render} formulæ in a human-readable way but not make them executable.  The Living Document \cite{GarciaEtAl:LivingDocument10} environment enables users to \textbf{annotate and share life science documents} and interlink them with Web knowledge bases, turning – like \sys – every single paper into a portal for exploring the underlying network.  However, life science knowledge structures, \eg proteins and genes, are relatively flat, compared to the tree-like and context-sensitive formulæ of \stem.  State-of-the-art \textbf{math e-learning systems}, including ActiveMath~\cite{activemath:on} and MathDox~\cite{mathdox:on}, also make papers executable.  However, they do not preserve the semantic structure of these papers in their human-readable output, which makes it harder for developers to embed additional services into papers.

\section{Conclusion and Outlook}
\label{sec:conclusion}

\sys makes documents executable on top of a content commons backed by structural ontologies.  Apart from mastering semantic markup – which we alleviate with dedicated editing and transformation technology – document authors, as well as authors of structural ontologies, only need expertise in their own domain.  In particular, no system level programming is necessary:  The semantic representations act as a high-level conceptual interface between content authors and the system and service developers.  Even developers can realize considerably new services as a client-side script that runs a query against the content commons.  This separation of concerns ensures a long-term compatibility of the knowledge hosted in a \sys instance with future demands.

\printbibliography
\end{document}